\documentclass[10pt,aps,prl,twocolumn,superscriptaddress]{revtex4}
\usepackage[linktocpage=true, colorlinks=true, urlcolor=blue, linkcolor=blue, citecolor=blue]{hyperref}
\usepackage{graphicx}
\usepackage{amsmath,amssymb}
\usepackage{color}
\usepackage{url}

\graphicspath{{Images/}}

\begin{document}

\title{
Off-axis digital holography with multiplexed volume Bragg gratings
}

\author{Leo Puyo}
\author{Jean-Pierre Huignard}
\author{Michael Atlan}

\affiliation{Centre National de la Recherche Scientifique (CNRS) UMR 7587, Institut Langevin. Paris Sciences et Lettres (PSL) Research University. Fondation Pierre-Gilles de Gennes, Institut National de la Sant\'e et de la Recherche M\'edicale (INSERM) U 979, Universit\'e Pierre et Marie Curie (UPMC), Universit\'e Paris 7. \'Ecole Sup\'erieure de Physique et de Chimie Industrielles ESPCI Paris - 1 rue Jussieu. 75005 Paris. France}

\date{\today}

\begin{abstract}

We report on an optical imaging design based on common-path off-axis digital holography, using a multiplexed volume Bragg grating. In the reported method, a reference optical wave is made by deflection and spatial filtering through a volume Bragg grating. This design has several advantages including simplicity, stability and robustness against misalignment.

\end{abstract}

\maketitle

\section{Introduction}

The success of the Zernike phase-contrast microscope~\cite{Zernike1942} has led to a number of subsequent phase-imaging methods, including Nomarski differential interference contrast microscopy~\cite{Lang1968} and Hoffman modulation contrast microscopy~\cite{HoffmanGross1975}. Those phase-contrast methods merge brightness and phase information in the recorded image. Later on, quantitative phase shifts of an object wave beating against a reference wave were measured by phase-contrast digital holography~\cite{CucheBevilacqua1999}. In digital holography, this phase is derived from the measurement of spatiotemporal variations in intensity of the interference pattern, digitized by a sensor array, and digital image rendering by wave propagation computation. Since then, holographic phase microscopy has emerged as a powerful imaging technique for probing quantitatively refractive index changes in transparent samples~\cite{MarquetRappaz2005}. Optical phase microscopy techniques based on interferometric recordings and computational image rendering have become widely available since then~\cite{MarquetDepeursingeMagistretti2014, McLeodOzcan2016, MaRajshekhar2016}, and have opened the way to tomographic phase microscopy \cite{Wolf1969, CharriereMarian2006, HaeberleBelkebir2010, CotteToy2013, JinZhou2017}.\\

On-axis interferometry, the original Gabor configuration for hologram recording~\cite{Gabor1948}, has the advantage of being easilly realized experimentally, but lacks of sensitivity~\cite{BhardwajlPanigrahi2016} and might prevent accurate phase reconstructions~\citep{Qu2010, Jericho2012}. In contrast, in off-axis recording configuration~\cite{LeithUpatnieks1962}, the average propagation directions of the reference and the object optical waves, interfering in the sensor plane, are slightly tilted. The spatial spectrum of the hologram is shifted by a quantity which scales as the average tilt angle. This tilt permits the separation and discrimination of self-beating and cross-beating interferometric contributions of the object and the reference optical fields; spurious interferometric contributions can be filtered-off~\cite{Cuche2000}, and phase imaging can be performed straightforwardly~\cite{CucheBevilacqua1999}. In addition, the optical pathes of light impinging on the sensor can have either both probed the sample in common-path configurations or taken different pathways in separate arms configurations. Common-path interferometry configurations, for which both waves travel the same pathways (quasi-)through the sample~\cite{Weijuan2009, Grishin2016, SerabynLiewerLindensmith2016, RostykusMoser2017} prevent phase noise from pathlength fluctuations of the reference beam decorrelated from the object beam, mode hops, and increase the overall stability of the interferogram. A significant improvement of off-axis interferometry with short coherence radiation was achieved by the use of diffraction gratings to tilt the coherence plane with respect to the direction of propagation of the reference wave, so that cross-interference patterns cover all the detector array~\cite{Monemhaghdoust2011, ChoiYang2012, DuboisYourassowsky2012, Witte2012, MonemahghdoustMonfort2013, SlabyKolman2013, GuoWangHu2017}.\\

\begin{figure}[h]
\centering
\includegraphics[width = 8.0 cm]{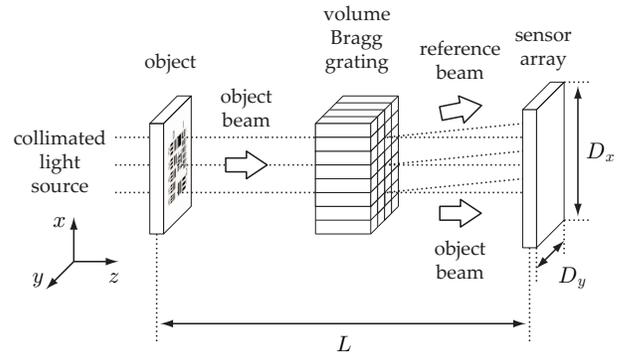}
\caption{Sketch of the experimental setup. A phase object is illuminated in transmission configuration, and the output field is partly deflected and filtered by a multiplexed volume Bragg grating to create an off-axis reference beam interfering with the object beam on a sensor array.}
\label{fig_Setup}
\end{figure} 
%

We propose here a new simple off-axis digital holographic microscopic scheme inspired by 1- point-diffraction interferometry approaches \cite{Smartt1975, MedeckiTejnilGoldberg1996, GaoYaoMin2011}, and their refinements~\cite{IndebetouwKlysubun1999, PopescuIkedaDasariFeld2006, WangMillet2011, Shaked2012}, in which a reference optical wave is formed by spatial filtering of a replica of the object wave. 2- off-axis implementations of phase-contrast digital holographic microscopy~\cite{CucheBevilacqua1999, GirshovitzShaked2013, LuLiuLau2014, Grishin2016, WallaceRider2015}, 3- lateral shearing interferometry~\cite{Primot1993, BonMaucort2009}, 4- the angular filtering properties of volume Bragg gratings~\cite{Kogelnik1969, Ludman1981}, recorded in the volume of doped glasses with photothermorefractive processes~\cite{EfimovGlebov1999}, and 5- an implementation of volume Bragg gratings for obtaining digital holograms in a self-reference configuration in a conventional microscope~\cite{MonemhaghdoustMontfort2014}. In the reported design, a reference optical wave, suitable for off-axis holographic interferometry is made by deflection through a volume Bragg grating.

\begin{figure}[h]
\centering
\includegraphics[width = 8.0 cm]{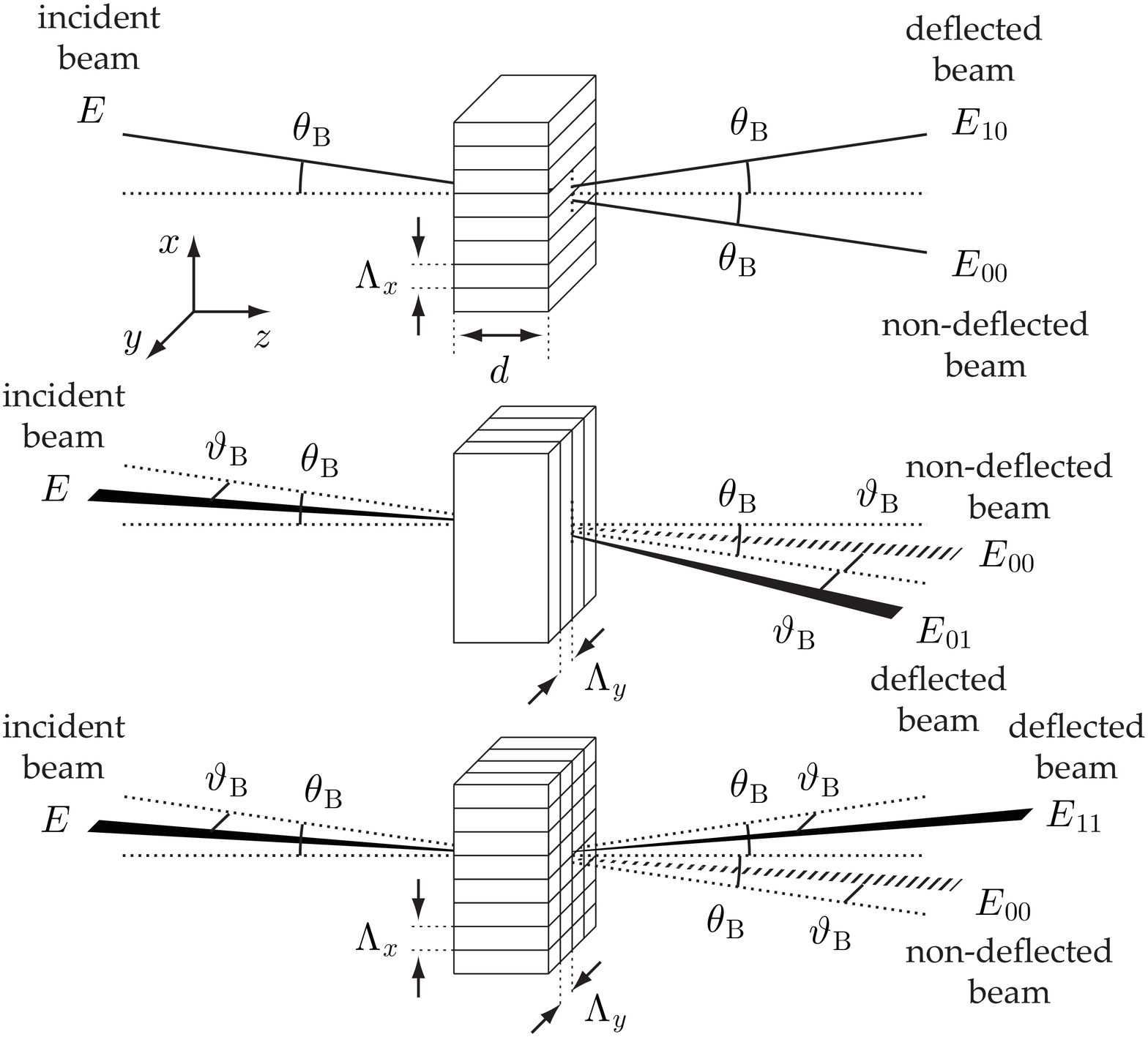}
\caption{Deflection and filtering of the optical field $E$ by volume Bragg gratings. The light deflected by $2\theta_{\rm B} \sim \lambda/\Lambda_x$ and/or $2\vartheta_{\rm B} \sim \lambda/\Lambda_y$ is filtered spatially. The $k_x$ content of the output field $E_{10}$ is lowpass filtered with an angular selectivity $\Delta\theta_1 \sim n_0\Lambda_x/d$. The $k_y$ content of the output field $E_{01}$ is lowpass filtered with an angular selectivity $\Delta\vartheta_1 \sim n_0\Lambda_y/d$. The $(k_x,k_y)$ content of the output field $E_{11}$ is lowpass filtered with an angular selectivity $(\Delta\theta_1,\Delta\vartheta_1)$. The non-deflected beam $E_{00}$ has an angular bandwidth $(\Delta\theta_0, \Delta\vartheta_0)$; it is not filtered.}
\label{fig_FilteringConfiguration}
\end{figure} 
\begin{figure}[h]
\centering
\includegraphics[width = 8.0 cm]{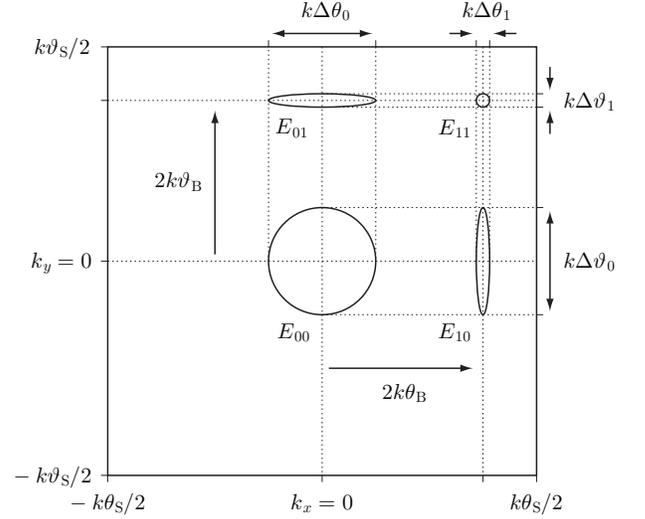}
\caption{Angular wavevector $(k_x,k_y)$ representation of the transmitted beams, in the Fourier plane of the exit face of the volume Bragg gratings. The angular bandwidth $(\Delta\theta_0, \Delta\vartheta_0)$ of non-deflected optical field $E_{00}$ is limited by the pupil and measured between the marginal rays touching its edges. The boundaries of the angular field of the detection are the Nyquist angles $(\pm\theta_{\rm S}/2, \pm\vartheta_{\rm S}/2)$, and its extent is the angular acceptance of the coherent detection $(\theta_{\rm S}, \vartheta_{\rm S})$. The shift and angular bandwidth of the fields $E_{10}$ and $E_{01}$ deflected by the first and second regular gratings of Fig.~\ref{fig_FilteringConfiguration} are represented. Those fields are allowed to pass when one of the Bragg conditions is not met with the multiplexed grating (Fig.~\ref{fig_FilteringConfiguration}, bottom)
but their presence is cancelled when both Bragg conditions are fulfilled. In that case, the reference field $E_{11}$ is deflected by $(2\theta_{\rm B},2\vartheta_{\rm B})$ with respect to the object beam, and low-pass filtered within the angular width $(\Delta\theta_1, \Delta\vartheta_1)$ of the Bragg diffraction efficiency function~\cite{Goodman2005}.}
\label{fig_DiffractionOrders}
\end{figure} 

\section{Experimental setup}

\subsection{Optical configuration}

In the experimental setup sketched in Fig.~\ref{fig_Setup}, a microscopic object is illuminated in transmission by a laser (Crystalaser DL660-100, wavelength $\lambda\sim$660 nm, coherence length 0.3 mm). The transmitted object field can be magnified by a microscope objective or merely occulted partially by a pupil, before passing through a diffractive optical element, and impinges on the sensor array of a camera (Ximea MQ042MG-CM, array size: 2048$\times$2048 pixels, pitch: $5.5 \, \mu \rm m$). In the absence of microscope objective, the angular bandwidth $(\Delta \theta_0, \Delta \vartheta_0)$ of the object field $E_{00}$ impinging on the sensor array is limited by the aperture stop of the system, which can either be a pupil introduced between the object and the sensor, or the sensor itself. In the former case, $\Delta \theta_0 \approx {a_x}/{l}$, and $\Delta \vartheta_0 \approx {a_y}/{l}$, where $a_x$ and $a_y$ are the aperture dimensions in the $x$ and $y$ directions, and $l$ is the object-to-pupil distance. In the latter case, $\Delta \theta_0 \approx {D_x}/{L}$, and $\Delta \vartheta_0 \approx {D_y}/{L}$, where $D_x$ and $D_y$ are the sensor dimensions in the $x$ and $y$ directions, and $L$ is the object-to-sensor distance. If a microscope objective is present, the angular bandwidth of the object field is limited by the numerical aperture ${\rm NA}$ of the objective : $\Delta \theta_0 = \Delta \vartheta_0 = 2 \arcsin({\rm NA})$. A usual method to engineer a reference optical wave suitable for off-axis holographic interferometry from the object wave itself consists in reducing the spatial frequency content of the latter by low-pass filtering of the transverse projections of the transmitted angular wavenumbers $(k_x,k_y)$ with a pinhole~\cite{Smartt1975, IndebetouwKlysubun1999, IndebetouwKlysubun2001, KlysubunIndebetouw2001}, or by defocus by curved mirrors~\cite{HongKim2013}, a spatial light modulator~\cite{RosenBrooker2007}, or lenses~\cite{NaikPedrini2014, MuhammadNguyen2016, SerabynLiewerLindensmith2016}. 

\subsection{Angular filtering by a multiplexed Bragg grating}

In our novel approach, a reference optical wave is made by deflection at the Bragg angle through a multiplexed grating. For this purpose, a transmission volume Bragg grating created by double interferometric exposure in a photothermorefractive glass was realized by OptiGrate~\cite{EfimovGlebov1999, CiapurinGlebov2006}. It displays a periodic change in the refractive index $n$ of the form 
\begin{equation}\label{eq_RefractiveIndex}
n(x,y) = n_0 + n_1 \sin \left( \frac{2 \pi x}{\Lambda_x} \right) + n_2 \sin \left( \frac{2 \pi y}{\Lambda_y} \right)
\end{equation}
where $n_0 = 1.498$ is the average refractive index of the glass, and $n_1 = n_2 \sim 1.4\times 10^{-5}$ is its modulation depth in each transverse direction, $x$, and $y$. It acts as a superposition of two perpendicular phase gratings, sketched in Fig.~\ref{fig_FilteringConfiguration}. The thickness of the multiplexed grating is $d = 8.9 \, \rm mm$. The grating periods are $\Lambda_x = 18.9 \, \mu \rm m$, and $\Lambda_y = 18.3 \, \mu \rm m$. The relative diffraction efficiency (for plane waves, when the Bragg condition is fulfilled) of the horizontal and the vertical grating are $|E_{01}|^2/(|E_{00}|^2+|E_{01}|^2) = 0.56$, and $|E_{10}|^2/(|E_{00}|^2+|E_{10}|^2) = 0.55$.\\

The Bragg condition defines the angle of deflection of a given input beam. For an unslanted grating, in the configuration sketched in Fig.~\ref{fig_FilteringConfiguration}, we have
\begin{equation}\label{eq_BraggAngle}
2 \sin( \theta_{\rm B} ) = \frac{\lambda}{\Lambda_x} \quad {\rm and} \quad 2 \sin( \vartheta_{\rm B} ) = \frac{\lambda}{\Lambda_y}
\end{equation}
where $\Lambda_x$ and $\Lambda_y$ are the grating periods. The spatial filtering properties of transmission volume Bragg gratings constrain the degree of collimation of the input beam allowed to be deflected by twice the Bragg angles. The angular efficiency, defined by the grating angular diffraction efficiency ~\cite{Goodman2005} sets the angular support $\Delta\theta_1$ and $\Delta\vartheta_1$ of the deflected fields in $x$ and $y$ respectively, given by 
\begin{equation}\label{eq_AngularEfficiency}
\Delta \theta_1 \approx n_0 \frac{\Lambda_x}{d} \quad {\rm and} \quad \Delta \vartheta_1 \approx n_0 \frac{\Lambda_y}{d}
\end{equation}
where $n_0$ is the average refractive index of the material, and $d$ is the thickness of the grating~\cite{Kogelnik1969, Ludman1981}. When both Bragg conditions are fulfilled for the multiplexed Bragg grating (Fig.~\ref{fig_FilteringConfiguration}(c)), the input angles of the object beam are $\theta_{\rm B}$ and $\vartheta_{\rm B}$, the deviation angles of the first diffraction orders are predicted by Bragg's law (Eq.~\ref{eq_BraggAngle}), and only low spatial frequencies $k_x \in [-k\Delta\theta_1/2,k\Delta\theta_1/2]$ and $k_y \in [-k\Delta\vartheta_1/2,k\Delta\vartheta_1/2]$ are allowed to pass in the deflected beam, yet the spatial frequency content of the non-deflected beam is unaffected by transmission through the grating. The angular representation of the corresponding optical fields, $E_{00}$ and $E_{11}$ respectively, are reported in Fig~\ref{fig_DiffractionOrders}. The shift and angular bandwidth of the fields $E_{10}$ and $E_{01}$ deflected by the first and second regular gratings of Fig.~\ref{fig_FilteringConfiguration} are also reported in Fig~\ref{fig_DiffractionOrders}. The spatial filtering property is the key to the realization of the tilted reference wave in common-path transmission  interferometric configuration. With the chosen grating thickness of $d \simeq 9 {\,\rm mm}$, this angular filter of acceptance $\Delta\theta_1=\Delta\vartheta_1\simeq 3.3 \,{\rm mrad}$ is equivalent to the angular selectivity $\sim D/f \simeq 3.3 \,{\rm mrad}$ of a pinhole of diameter $D = 330 \,\mu {\rm m}$, set in the focal plane of a converging lens of focal length $f = 10 \,{\rm cm}$ in a point-diffraction interferometer~\cite{Smartt1975, MedeckiTejnilGoldberg1996, GaoYaoMin2011}. The choice of grating periods $\Lambda_x$ and $\Lambda_y$ is dictated by the Nyquist–Shannon sampling theorem : the deflection angles of the propagation directions of $E_{11}$ and $E_{00}$ have to satisfy $2\theta_{\rm B} \in [-\theta_{\rm S}/2,\theta_{\rm S}/2]$, and $2\vartheta_{\rm B} \in [-\vartheta_{\rm S}/2,\vartheta_{\rm S}/2]$ where
\begin{equation}\label{eq_AntennaAcceptanceAngle}
\theta_{\rm S} \approx \frac{\lambda}{d_x} \quad {\rm and} \quad \vartheta_{\rm S} \approx \frac{\lambda}{d_y}
\end{equation}
are the angular acceptances of the coherent detection on an array detector~\cite{Siegman1966}, with pixels spaced by $d_{x}$ and $d_{y}$ along $x$ and $y$, respectively. The pixel pitches of the camera used in the experiments are $d_{x} = d_{y} = 5.5 \, \mu \rm m$. Eq.~\ref{eq_BraggAngle}, Eq.~\ref{eq_AngularEfficiency}, and Eq~\ref{eq_AntennaAcceptanceAngle} are valid for small input and output angles, in the paraxial approximation~\cite{Goodman2005}. In order to achieve typical off-axis detunings between the object and the reference waves, deflection angles have to be about one half of the Nyquist angles $\theta_{\rm S}/2$ and $\vartheta_{\rm S}/2$ typically, and hence satisfy the relations $2\theta_{\rm B} \sim \theta_{\rm S}/4$, and $2\vartheta_{\rm B} \sim \vartheta_{\rm S}/4$, which impose grating periods of $\Lambda_x \sim 4 d_x$ and $\Lambda_y \sim 4 d_y$.

\begin{figure}[h]
\centering
\includegraphics[width = 8.0 cm]{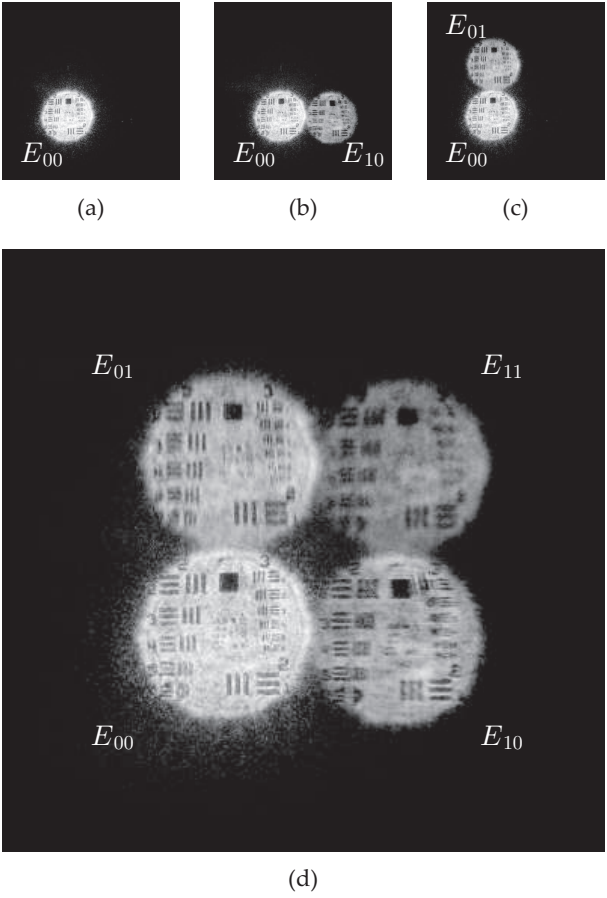}
\caption{Experimental images of the target recorded for different tilt angles of the thick Bragg grating, obtained with a physical lens. Bragg conditions (Eqs.\ref{eq_BraggAngle}) not satisfied (a), satisfied only for $\theta$ (b), satisfied only for $\vartheta$ (c), satisfied for both $\theta$ and $\vartheta$ (d).}
\label{fig_ExperimentalBraggOrdersLaser658nmImgFullTarget}
\end{figure} 

\subsection{Experimental validation of angular filtering and deflection by a multiplexed Bragg grating}

To illustrate the spatial filtering properties of the Bragg orders, we placed a converging lens between the grating and the sensor array to form the image of the resolution target in the detection plane. In Fig.~\ref{fig_ExperimentalBraggOrdersLaser658nmImgFullTarget}(a), the Bragg condition is not met for any beam direction, only the non-deflected beam (field $E_{00}$) is transmitted. The angular bandwidth of the object field $E_{00}$ is completely transmitted through the grating, and spatial frequencies are bounded by $(k\Delta\theta_0, k\Delta\vartheta_0)$, as sketched in Fig.~\ref{fig_DiffractionOrders}. In Fig.~\ref{fig_ExperimentalBraggOrdersLaser658nmImgFullTarget}(b), the Bragg condition is met only for $\theta$ tilt ($x$ direction); the non-deflected beam (field $E_{00}$) and the beam deflected by $2\theta_{\rm B}$ (field $E_{10}$) are transmitted. The angular bandwidth of the deflected field $E_{10}$ is reduced. Its spatial frequencies are bounded by $(k\Delta\theta_1, k\Delta\vartheta_0)$, as sketched in Fig.~\ref{fig_DiffractionOrders}. Hence, horizontal bars of the deflected image are no longer visible. In Fig.~\ref{fig_ExperimentalBraggOrdersLaser658nmImgFullTarget}(c), the Bragg condition is met only for $\vartheta$ tilt ($y$ direction); the non-deflected beam (field $E_{00}$) and the beam deflected by $2\vartheta_{\rm B}$ (field $E_{01}$) are transmitted. 
The angular bandwidth of the deflected field $E_{01}$ is reduced. Its spatial frequencies are bounded by $(k\Delta\theta_0, k\Delta\vartheta_1)$, as sketched in Fig.~\ref{fig_DiffractionOrders}. Hence, vertical bars of the deflected image are no longer visible. When the Bragg condition is fulfilled for both directions, the non-deflected beam $E_{00}$, the deflected and spatially-filtered beams $E_{10}$ and $E_{01}$ are transmitted, as reported in Fig.~\ref{fig_ExperimentalBraggOrdersLaser658nmImgFullTarget}(d). A fourth beam, deflected both in $\theta$ and $\vartheta$ by $2\theta_{\rm B}$ and $2\vartheta_{\rm B}$ respectively is transmitted (field $E_{11}$). The angular bandwidth of the deflected field $E_{11}$ is reduced. Its spatial frequencies are bounded by $(k\Delta\theta_1, k\Delta\vartheta_1)$, as sketched in Fig.~\ref{fig_DiffractionOrders}. Hence, both horizontal and vertical bars of the deflected image are no longer visible.

\section{Off-axis hologram reconstruction}

\subsection{Fresnel transformation}

Holograms were reconstructed by Fresnel transformation~\cite{Goodman1967, SchnarsJuptner1994, Schnars2002, KimYuMann2006, PicartLeval2008, VerrierAtlan2011, PiedrahitaQuintero2015, Latychevskaia2015}; in practice, we used the software Holovibes~\cite{Holovibes} for real-time hologram rendering, which performed image rendering of complex-valued holograms $H(x,y,t)$ from the stream of rescaled interferograms $I(x,y,t)$
\begin{eqnarray}\label{eq_FresnelTransform}
\nonumber H(x,y,t) = \frac{i}{\lambda z}\exp \left( -ikz \right) \iint I(x',y',t)\\
\times \exp \left[\frac{-i \pi}{\lambda z} \left((x-x')^2 + (y-y')^2\right) \right] {\rm d}x' {\rm d}y'
\end{eqnarray}
The parameter $z$ is the reconstruction distance for which the image of the target appears on the magnitude of the hologram $|H|$. The interferometric contributions in the interferogram plane that create fringe sets  (Fig.~\ref{fig_InterferogramAndHologramLaser658nm}(a), and Fig.~\ref{fig_InterferogramAndHologramLaser658nm}(b)) are separated in the Fourier reciprocal plane (Fig.~\ref{fig_InterferogramAndHologramLaser658nm}(c)) and the hologram reconstruction plane (Fig.~\ref{fig_InterferogramAndHologramLaser658nm}(e)).

\subsection{Amplitude imaging}

\begin{figure}[h!]
\centering
\includegraphics[width = 8.0 cm]{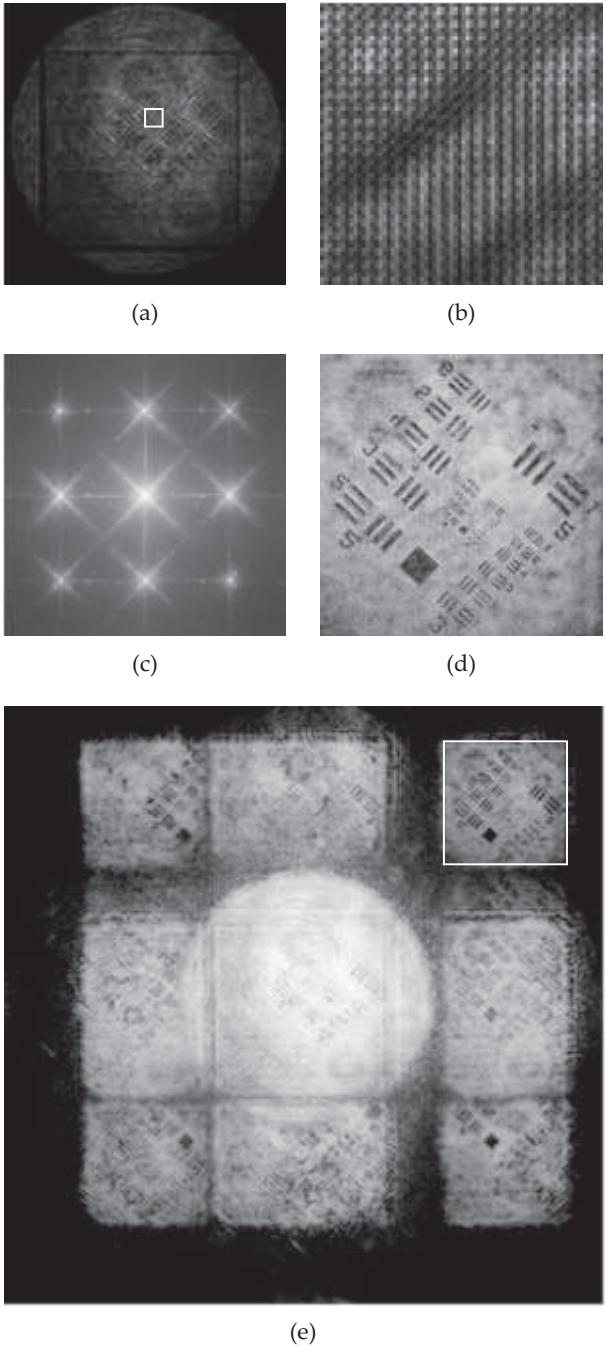}
\caption{Amplitude contrast target. (a) Recorded interferogram $I$, where the shadow delimiting the edges of the diffraction grating is cast. (b) Magnified view of the fringes in the highlighted region in (a). (c) Amplitude distribution in the Fourier plane $|{\cal F} \{ I \}|$ (log. scale). (d) Magnified view of the off-axis area of the magnitude of the hologram $|H|$ displayed in (e). Amplitude distribution of the Fresnel transform of the interferogram (a), at the distance 0.216 m (e).}
\label{fig_InterferogramAndHologramLaser658nm}
\end{figure} 

\begin{figure}[h]
\centering
\includegraphics[width = 8.0 cm]{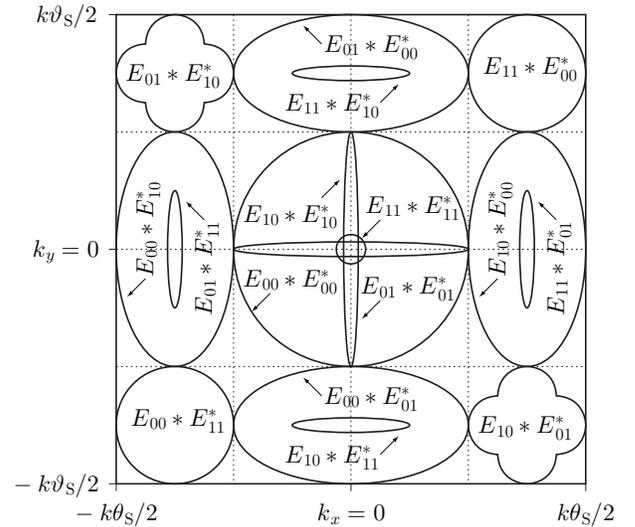}
\caption{Sketch of the 16 interference terms locations in the Fourier plane of the interferogram.}
\label{fig_InterferometricContributions}
\end{figure}

In order to record holographic interferograms, we removed the lens between the grating and the sensor array, and we increased the aperture stop of the system to widen the lateral extension of the deflected fields, and make the four contributions interfere in an overlapping region covering most of the sensor area, as shown in the experimental interferogram displayed in Fig.~\ref{fig_InterferogramAndHologramLaser658nm}(a). A zoomed view in the region of interest depicted by the box, showing the juxtaposition of horizontal and vertical fringes, is reported in Fig.~\ref{fig_InterferogramAndHologramLaser658nm}(b). When the Bragg conditions are fulfilled for both directions, angular spectra of $E_{10}$, $E_{01}$, and $E_{11}$ are filtered by the Bragg grating angular selectivity, as sketched in Fig.~\ref{fig_DiffractionOrders}. The total transmitted field is the sum of four components forming an interferogram on the sensor array $I = |E_{\rm t}|^2$, where 
\begin{equation}\label{eq_SumOfFields}
E_{\rm t} = E_{00} + E_{10} + E_{01} + E_{11}
\end{equation}
Each set of fringes corresponds to the interference between couples of field components, except self-beating contributions. The reconstructed hologram $H$ by linear Fresnel transformation (Eq.~\ref{eq_FresnelTransform}) shifts those contributions according to this fringe structure. The spatial Fourier transform ${\cal F} \{ I \}$ of the interferogram $I$ displayed in Fig.~\ref{fig_InterferogramAndHologramLaser658nm}(a) takes the form 
\begin{equation}\label{eq_SpatialConvolutionFields}
{\cal F} \{ I \} = E_{\rm t} * E_{\rm t}^{*}
\end{equation}
where $*$ is the spatial convolution product. For the sake of notation simplicity, the fields $E$ either refer to distributions in the sensor plane, or at the exit face of the volume Bragg grating, or their reciprocal planes. The magnitude $|{\cal F} \{ I \}|$ is displayed in logarithmic scale in Fig.~\ref{fig_InterferogramAndHologramLaser658nm}(c), on which one can see nine diffraction locations of the 16 interferometric terms. A sketch of the 16 interference terms locations in a reciprocal plane of the interferogram is reported in Fig.~\ref{fig_InterferometricContributions}.\\

Among the 16 interferometric terms of ${\cal F} \{ I \}$, the term $E_{00} * E_{11}^{*}$ highlighted in Fig.~\ref{fig_InterferogramAndHologramLaser658nm}(e), and displayed in Fig.~\ref{fig_InterferogramAndHologramLaser658nm}(d), and its complex conjugate $E_{00}^{*} * E_{11}$ are shifted in opposite corners of the reconstructed hologram, and used as off-axis holograms. These contributions are the result of the interference between the transmitted field $E_{00}$, and the field of reduced angular support in both directions $(x,y)$ $E_{11}$ which acts as a flattened reference wave. In the remaining corners, the beating contributions $E_{10} * E_{01}^{*}$ and $E_{10}^{*} * E_{01}$ are present. These terms are the result of the beat between transmitted and partly filtered fields in the horizontal and vertical directions. In the four sides, the beating contributions $E_{10} * E_{01}^{*}$ and $E_{10}^{*} * E_{01}$ are present. These terms are the result of the beat between transmitted and partly filtered fields in the horizontal and vertical directions.

\begin{figure}[h!]
\centering
\includegraphics[width = 8.0 cm]{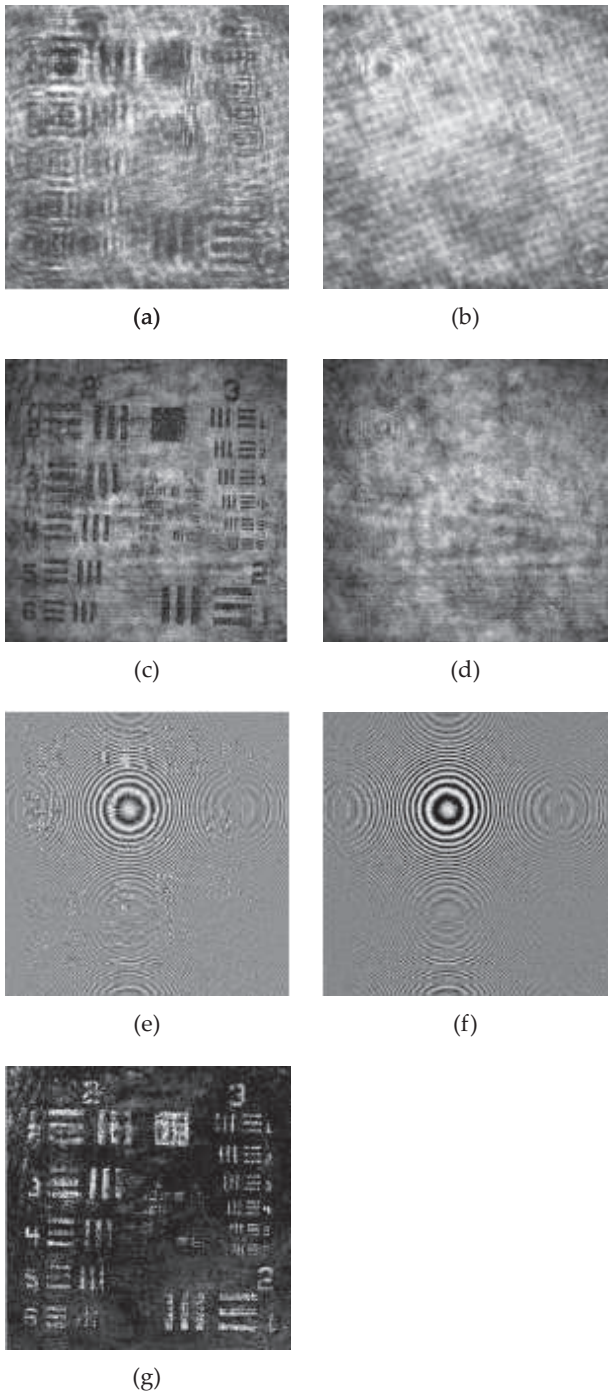}
\caption{Image rendering process. Central region of the interferogram acquired, with (a) and without target (b). Amplitude of the holograms reconstructed in the object plane, with (c) and without target (d). Phase of the holograms reconstructed in the object plane, with (e) and without target (f). (g) Difference of the phase holograms (e) and (f).}
\label{fig_PhaseAndAmplitudeHologramRendering}
\end{figure} 

\subsection{Phase imaging}

In order to obtain phase images, holograms were processed to remove artifacts generated by the off-axis configuration~\cite{TrujilloCastaneda2016} and the phase curvature~\cite{KreuzerJericho2001} of the reference wave. The interferometric order $E_{00} * E_{11}^{*}$ was selected and re-centered in the Fourier space while the rest of the Fourier components were cropped out~\cite{CucheBevilacqua1999, TrujilloCastaneda2016}. These operations allowed for the removal of signal corresponding to other interferometric contribution and also of the off-axis phase tilt. A Fresnel transformation (Eq.~\ref{eq_FresnelTransform}) was carried out onto the interferograms acquired in the presence [Fig.~\ref{fig_PhaseAndAmplitudeHologramRendering}(a)] and absence [Fig.~\ref{fig_PhaseAndAmplitudeHologramRendering}(b)] of target. The phase image reported in Fig.~\ref{fig_PhaseAndAmplitudeHologramRendering}(g) is the difference between the phase of the reconstructed holograms acquired with [Fig.~\ref{fig_PhaseAndAmplitudeHologramRendering}(e)] and without [Fig.~\ref{fig_PhaseAndAmplitudeHologramRendering}(f)] target. The accuracy of the phase measurement might be hindered by a not perfectly flat reference wave in the reported results.

\section{Conclusions}

In conclusion, the reported off-axis common-path holographic interferometer design performs spatial filtering with a multiplexed volume Bragg grating, which makes it suited to the detection of optical absorption and index variations. The described interferometer is lensless, and may be adapted to microscopic imaging. Lateral resolution, robustness against vibration and aberration, spatiotemporal coherence requirements of the radiation and accuracy of phase imaging might be further investigated. The reported results made use of one thick grating; alternatively, the use of two thinner multiplexed gratings could result in the same deflection and filtering properties~\cite{OttSeGall2015}.

\section*{Funding}

This work was supported by the Investments for the Future program (LabEx WIFI: ANR-10-LABX-24, ANR-10-IDEX-
0001-02 PSL*), and European Research Council (ERC
Synergy HELMHOLTZ, $\#$610110).

\end{document}